# Enabling Embodied Analogies in Intelligent Music Systems

Fabio Paolizzo, Ph.D.
University of California Irvine, Department of Cognitive Sciences and Department of Dance.
University of Rome Tor Vergata, Department of Electronic Engineering.

When human beings make or listen to music, they regularly draw on analogies to other forms of knowledge and embody how they feel through their own body. Research on interactive music systems has gradually shifted from "dialogue" between human and computer to information retrieval, thus posing the risk of commodifying and reifying the encounter with technology [1]. On the contrary, recent advances brought a dramatic increase in human-centered and reflexive retrieval and indexing methods based on subjective concepts such as emotion, preference and aesthetics [2]. Interestingly, recent studies on music cognition suggest that listeners are able to change their semantic relations with the sonic world through functional adaptation at the level of sensing, acting and coordinating between action and perception, in biological, psychological, and cultural terms that involve motor, kinesthetic, haptic and visual, besides the purely auditory components [3]. Implementing some understanding of how we feel and attribute meaning, when interacting with music technology, requires next-generation systems of interactive music, music intelligence and music information retrieval to operate in terms that can represent human cognition as multimodal and embodied.

In the literature of intelligent (and interactive) music systems, early approaches to mutual listening [1] [4] — a milestone in the field — improve with stylistic reinjection; in OMax, a symbolic model recombines the player's discourse, 'who is constantly confronted to a reinterpreted version of his own playing' [5]. Response is similarly contextual in PyOracle, which builds predictive oracles adaptively on arbitrary feature data without quantization, for machine improvisation to match a live musical input in a meaningful way for the user [6]. A step further is taken in interactive reflexive musical systems, which extend the research on active lexicons created from a knowledge model of the user's interaction [4], for the exploration and manipulation of style [7]. Other systems implement reflexivity by modelling extra-musical information to simulate intentionality in automatic music generation [8]. Still, the gap between computational creativity and human music making is evident, as contrarily to those systems, humans regularly draw on analogies to other forms of knowledge.

Regardless of the widespread interest in computational methods that leverage on the relevance of affect to cognition, current research has yet to give greater consideration to the relation between analogy and emotion. We can easily recall three kinds of emotional analogies, such as 'analogies and metaphors about emotions […], analogies that involve the transfer of emotions […], analogies that generate emotions' [9]. Furthermore, the study of analogical inference regarding emotions should at least take into some account notions of cognition as embodied [10] and the relation between language, visual sequencing and motor planning as information that is similarly structured sequentially [11]. Advancing research for automatic music generation requires enabling systems to classify, recognize and predict emotional states in experiences of 'musicking' [12] that comprise domains such as the sonic, visual, motion and linguistic. Investigating how we translate between these domains, "making analogies" is the aim of the Musical-Moods project: a mood-indexed database of scores, lyrics, musical excerpts, vector-based 3D animations, and dance video recordings [13].

Music emotion recognition is an emerging and cross-disciplinary field spanning information retrieval (audio, symbolic, metadata) and machine learning, on a strong backing of music cognition (semiology of music and psychology) and music theory. Strong performance was initially reported in tasks of music mood classification by using audio features only [14], and a number of algorithms are currently available (*i.e.*, MIRtoolbox [15]). In the last few years, other methods incorporating metadata have demonstrated the need for a multimodal approach with a strong focus on human language processing [16] [17]. Recent research has confirmed this by ranking linguistic features higher than audio features, in terms of valence, and by suggesting that symbolic features are useful only when language data cannot be used [18]. Mindprint-based classification is a state-of-the-art technique for language data, which captures sophisticated linguistic features involving semantic, syntactic, and valence information, and notably is not limited to moods leveraged by domain-specific knowledge of interest [19] (*i.e.*, Russell's [20] [21] or MIREX [8]). Each set of these features, namely a mindprint, is not targeted, as affect features for emotions would be, and rather spans intentions, attitudes, and emotions. In the study, mindprints are used to represent commonly investigated moods (*i.e.*, anger, annoyance, frustration, boredom, tiredness, calm, satisfaction, pleasantness, joyfulness, etc.), as well as other moods that those broader categories do not capture accurately but, nevertheless, can be expressed through music (*e.g.*, ecstasy, giddiness, craziness, mellowness, etc.).

Research in the Musical-Moods project encompasses a cross-modal approach to machine learning for validating the model and evaluating the results, through creative practice, in order to realize an innovative music mood classification using audio and metadata. In the last few years, collection of ground truth for mood labels has been increasingly carried out by using "games-with-a-purpose" [22] [23], as based on the "wisdom of the crowds" phenomenon [24] [25] [26] and thus less expensive than listener surveys [27] [28] or social tags [29]. Recently, the approach has been integrated in methods for multimodal distributed semantics, thus showing that 'the performance of computational models of meaning improves once meaning is grounded in perception' [30]. The present study proposes a methodology to validate the model through a taxonomy of relations between the musical, linguistic and motion domains. This taxonomy will provide some understanding of how humans "make analogies" in accordance to how they feel, for further exploitation from next-generation human-centered and reflexive, retrieval and indexing technology.

**Research methodology and approach**

Digital scores inclusive of lyrics will form an effective and large corpus for classification. Such scores are easily gathered from public domain collections of music, converted into audio files by using a software synthesizer, and integrated in a database. Datasets are being collected using a game-with-a-purpose, such as Word Sleuth game-with-a-purpose [31], for Internet users to validate the linguistic model. Furthermore, motion capture technology, including a Vicon system [32], are being used to track the dancers' counterpart of physical expression, in order to gather data and further annotations for realizing the cross-modal taxonomy [33] [34] [35] [36]. In a typical session with the motion capture system, a dancer is asked to: (i) improvise on a music file played from the database and to annotate the mood (*e.g.*, among a proposed set), which s/he considers as expressed by the music; (ii) annotate the mood expressed by the lyrics of another music file in the database; (iii) play the online annotation game. The database then integrates datasets from the motion capture system (*i.e.*, velocity, acceleration, and values of position), their vector-based 3D representation, video signals, and different scores, their audio conversion and lyrics files, for each mood in the model. Third-party computer programs (*i.e.*, MAX/MSP) can be used for carrying out the task. Ground truths and metrics of the human performance on the expression and identification of moods, in music, lyrics and dance, can thus be compared for identifying audio features (*i.e.*, notes per second, attack time, roughness, peak sound level, spectral contrast, MFCCs., CQT, etc.) with some account of symbolic features (*i.e.*, tempo, note duration, note density, etc.). The approach will allow for pre-processing of all the data into mood-indexed excerpts.

Further validation and refinement of the database will be achieved by the realization of music works with interactive and intelligent music systems. This way, case studies will support the identification of features that are best highlighted by music performance or composition practice (*e.g.*, nuances of musical interpretation, cross-domain and analogy-based usage of the datasets, extensive combination of datasets excerpts, superimposition of datasets layers, etc.). In outreach events, the audience will be offered the opportunity to vote for the most interesting works. Finally, a metric of the audience and artists' preference, so generated, will be integrated in the database.

**Originality and innovative aspects**

The area intersecting interactive and intelligent music systems, information retrieval, as well as computational creativity and music emotion research at large, is rapidly growing and expanding its boundaries, and there is scope for major developments in a short time. Notably, a mood-indexed, multimodal and open-data database, spanning audio, symbolic, linguistic, vector-based 3D animations, and dance video recordings, such as that the Musical-Moods project addresses, does not yet exist. In the Musical-Moods project, the database will be ready for further exploitation, thus opening-up new creative possibilities for different types of users, such as musicians, dancers, choreographers, 3D graphics animators, actors, video makers, and researchers operating within these and related areas (*i.e.*, image and video feature extraction and classification). The methodology discussed above is specifically aimed at systems working with various media in these areas. Furthermore, as the way in which humans process and attribute meaning relies on a cross-modal experience of the world, investigating those processes represents a major challenge to the progressive encroaching of technology in content creation, production, fruition and consumption. Thus, a more general objective is to evaluate the contribution that the adoption of elements from the present cross-modal methodology brings to the broader fields of computational creativity, music making and music emotion research.

**Funding Acknowledgement**

The research is supported by the EU through the MUSICAL-MOODS project funded by the Marie

Sklodowska-Curie Actions Individual Fellowships Global Fellowships (MSCA-IF-GF) of the Horizon 2020 Programme H2020/2014-2020, REA grant agreement n° 659434.